\documentclass{jpsj2}

\def\runauthor{Miyake, Pouroskii, Vildosola, Biermann, Georges}
\def\LAs{$\rm{LaFeAsO}$~}

\title{d- and f-orbital correlations in the REFeAsO compounds}

\author{
\textsc{T. Miyake}$^{1,4}$
\thanks{E-mail address: t-miyake@aist.go.jp},
\textsc{L. Pourovskii}$^{2}$
\thanks{E-mail address: leonid@cpht.polytechnique.fr},
\textsc{V. Vildosola}$^{2,3,4}$, \textsc{S.  Biermann}$^{1,3}$,
\textsc{A. Georges}$^{2,4}$ }

\inst{
$^{1}$ Research Institute for Computational Sciences, AIST, Tsukuba
305-8568, Japan\\
$^{2}$ Centre de Physique Th\'{e}orique, \'{E}cole Polytechnique, CNRS, 91128 Palaiseau, France\\
$^{3}$ Departamento de F\'{\i}sica, Comisi\'{o}n Nacional de Energ\'{\i}a At\'{o}mica (CNEA-CONICET),
San Mart\'{\i}n, Provincia de Buenos Aires, Argentina\\
$^{4}$ Japan Science and Technology Agency, CREST\\
}

\abst{%
We estimate theoretically the strength of the local Coulomb interaction for the Fe 3$d$ and Ce 4$f$ shells in the
REFeAsO compunds.
In LaFeAsO and CeFeAsO  we obtain values of the local Coulomb interaction parameter $U$ for both Fe and Ce
which are larger than those of elemental Fe and Ce metals.
The Fe 3$d$ bandwidth of REFeAsO is found to increase slightly as one moves along the RE-series.
Using a combined local density approximation and dynamical mean-field theory (LDA+DMFT) approach,
we study the behaviour of the localized 4$f$ states along the rare-earth oxyarsenides REFeAsO series
(RE=Ce,Pr,Nd). In CeFeAsO the occupied Ce 4$f$ band is located just below the Fe 3$d$ band
leading possibly to a Kondo screening of the 4$f$ local moment under applied pressure, while
the unscreened local moment behaviour is expected for the Pr and Nd compounds.
}

\kword{%
hight-Tc superconductors, Fe-oxipicnitides, heavy-fermions
}

\begin{document}
\maketitle

\section{Introduction}
The discovery of high-T$_c$ superconductivity in FeAs-based
materials \cite{Kamihara-LOFA} --
the first non cuprates high-Tc superconductors
-- has triggered intensive research lately.
As in the cuprates, the superconducting state (SC) occurs close
to a long range SDW AF-type order \cite{cruz}
and is believed to originate from the quasi 2D FeAs layers,
in which Fe forms a square lattice.
Both the FeAs-based superconductors\cite{struct,zhao} and the
cuprates\cite{slezak} display a strong sensitivity of T$_c$ on
the relative atomic positions within the transition metal layer.
In view of these similarities and of the well-known strongly
correlated nature of the Cu-$d$ band in
the high-Tc cuprates, the strength of local Coulomb correlations in the
FeAs superconductors is a key issue to be clarified.
On the one hand, LDA can, apart from some shifts, account for the photoemission
spectra qualitatively well \cite{malaeb,ding}.
Several works based on X-ray absorption experiments
are also pointing to a weakly correlated behavior of these compounds
\cite{photo} while a feature due to Fe 3$d$ states localization,
the lower Hubbard band, which is predicted by dynamical mean field theory (DMFT)\cite{haule}
in the intermediate to strong correlation regime, has not yet
been seen in spectroscopic experiments.
%
%
%
%
On the other hand, it has been established that LDA calculations overestimate the strength of the Fe-As bonding
leading to a substantial underestimation of the Fe-As bond length, this disagreement being only partially
 improved by GGA \cite{mazin2}. However, somewhat confusingly, LDA calculations (and not GGA)
 within optimized structures can account for the magnetic properties\cite{mazin2,pickett}.
Some experimental facts also point to strong correlations, such as the bad
metallicity (high resistivity) of the normal state\cite{Kamihara-LOFA,dong},
the absence of a Drude peak in the optical conductivity\cite{dong} and strong
temperature-driven spectral weight transfers in optical measurements\cite{boris}.

Hence, whether the Fe 3$d$ states in the oxypnictides are weakly or rather strongly correlated remains a
largely open question.
Another important issue is whether the localized RE 4$f$ states play any significant role in the
electronic properties of these materials, e.g. by hybridizing with the Fe 3$d$ states and
whether these RE states affect the superconducting properties of the REFeAsO compounds.

In this paper we study the evolution of the Fe 3$d$  bandwidth and estimate
theoretically the strength of the local Coulomb interaction for both the
Fe 3$d$ and RE 4$f$ states, within the constrained RPA and LDA methods, respectively.
We also study the evolution of the RE 4$f$ band along the REFeAsO series (RE=Ce,Pr,Nd)
using an approach that combines the local density approximation and
dynamical mean-field theory (LDA+DMFT).

\section{Band structure trends}

The evolution of $T_c$ along the series of rare earth
elements, and possible correlations with structural
properties have attracted particular interest.
While LaFeAsO$_{1-x}$F$_x$ has a $T_c$ of 26 K,
the critical temperature is drastically increased
by replacing La by other rare earth ions
(RE=Ce,Pr,Sm,Nd,Gd),
up to $\sim$ 55K for Sm .
Possible correlations between the evolution
of $T_c$  with  changes in the structural
parameters due to decreasing size of the rare-earth ions
along the series have been reported
in the literature\cite{ren3,zhao}.
Fig. 1 shows the evolution of the
Fe-bandwidth from La to Sm calculated using the
Full-Potential APW+local orbitals method as implemented in the Wien2k\cite{wien2k} code: the increase
of the bandwidth is due to the decrease of the Fe-Fe distance and to the
corresponding decrease of the Fe-As bond angle $\theta_3$.
The most natural candidate to search for strong
d-electron correlations thus appears to be the first
member of the series, LaFeAsO, which has the smallest bandwidth.

\begin{figure}[h]
\begin{centering}
\includegraphics[clip,scale=0.3]{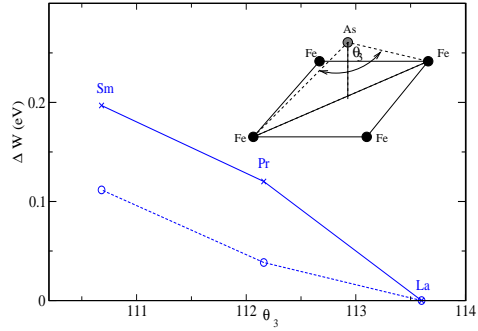}
\par\end{centering}
\caption{The change $\Delta W$ in the Fe 3$d$ full bandwidth $W$ (solid line) and the bandwidth of its occupied part $W_{occ}$ (dashed line)
  for several RFeAsO materials given with respect to $\theta_3$. The change $\Delta W$ is referenced to calculated value for \LAs at its experimental structure\cite{cruz} (4.25 eV for $W$ and 1.95 eV for $W_{occ}$). The structural data for Pr and Sm compounds were
taken from Ref. \cite{POFA} and \cite{SOFA}, respectively. {\it Inset}:Schematic Fe$_4$As pyramid for LaFeAsO in
which $\theta_3$ angle is indicated.}
\end{figure}

\section{d-orbital correlations}

We have calculated the Coulomb interactions
within the constrained RPA method proposed in Ref.\cite{aryasetiawan_prb04}.
Within the random phase approximation (RPA),
the polarisation P of a solid is calculated as
a sum over all possible transitions between
empty and occupied states, weighted by their
respective energy differences.
Whereas the fully screened Coulomb interaction
is then calculated by screening the bare Coulomb interaction
v by P, the onsite Hubbard interaction is screened
by a partial polarisation function that excludes
all those screening processes that are included
in the many-body model:
\begin{equation}
U = \frac{v}{1- P_r v}
\end{equation}
For a model retaining only d-bands, for example,
$P_r = P - P_d$, where $P_d$ is the polarisation
function corresponding to d-d transitions only.

Obviously, matrix elements of the local interaction depend
on which states are included in the model (and are
thus not allowed to screen U) and - in a more
trivial fashion - on the spread of the Wannier
functions used for taking the matrix elements.
We have calculated matrix elements of U and W
in maximally localised Wannier basis, corresponding
to different models:
(i) The first one is a ``d-only model'', where
only d-d transition were cut out, and the Wannier
construction was performed for the manifold of
d-states only.
(ii) The second choice is a ``dpp-model'', where
the whole block of Fe-d, As-p and O-p bands was
used for the Wannier construction and for the
truncation of the polarisation.
(iii) The third option is a hybrid between the
first two, in the sense that only the d-d transitions
were cut out in the polarisation, but the Wannier
functions from the dpp-model were used.
When performing many-body calculations with a hamiltonian containing all dpp-states
but including only U for the d-states, this last procedure should
be the most appropriate way of evaluating the proper value of U
to be used in such a framework.

\begin{table}
\begin{tabular}{|c|c|c|c|c|}
\hline
 & v &  U & J \tabularnewline
\hline
\hline
\emph{d} & 15.99 & 2.92 &  0.43 \tabularnewline
\hline
\emph{dpp} & 20.31 & 4.83 &  0.61 \tabularnewline
\hline
\emph{d-dpp} & 20.31 & 3.69  & 0.58\tabularnewline
\hline
\end{tabular}
\caption{Bare and partially screened
Coulomb interactions v and U (orbitally averaged), as well as the Hund's
coupling for LaFeAsO in the different Wannier constructions
described in the text.}
\end{table}

As expected, the Hubbard interaction is smallest for
the d-only model, where most of the screening has already
been taken into account before the many-body treatment,
and the Wannier orbitals are very extended \cite{vildosola}.
The comparison between the dpp and d-dpp models allows
to identify the contributions from the two effects:
the screening processes taken into account are the
same as in the d-model, but the orbitals are those of
the dpp-model.

In absolute values, our results are somewhat bigger
than the ones obtained in Ref.\cite{nakamura,footnote},%
and indicate that within a d-dpp model U values beyond
4 eV are not unrealistic.
According to \cite{haule}
this would place LaFeAsO close to the Mott transition.
As discussed above, experimental photoemission spectra may
not support this view. It is possible that effects not included in the
single-site DMFT calculations (non-local Coulomb interactions mediated
e.g. by the strong Fe-As hybridization, non-local
self-energy effects, ...) are important to settle this issue.


%

\section{f-orbital correlations}

In order to study the evolution of the strongly localized 4$f$ states
along the rare-earth oxyarsenides REFeAsO series (RE=Ce,Pr,Nd) and their interaction with
other bands, we have performed {\it ab initio} calculations
using the combined local density approximation
and dynamical mean-field theory (LDA+DMFT) approach, in the
fully-selfconsistent framework described in \cite{pour2007}.
The local self-energy of the $4f$ shell has been computed by employing the
atomic (Hubbard-I) approximation\cite{hub1}.
This approach to local correlations
has been shown to be appropriate for the localized 4$f$ shells of
rare-earths compounds \cite{pour2007}.
We have taken values of of the Slater integrals $F^2$, $F^4$,
 and  $F^6$ for the local Coulomb interaction from the optical measurements of \cite{carnal}.
 In order to determine the value of the local Coulomb interaction $U$ on the 4$f$ shell,
 we have performed constrained LDA calculations for CeFeAsO.
 and obtained a value of about $9.7$~eV for $U$ on the Ce 4$f$ shell.
 This is substantially larger than the usual range of $U$ values for pure Ce.
 This can be explained by the quasi two-dimensional environment of the rare-earth sites
 in the case of oxypnictides, resulting in a rather poor screening of the local
 Coulomb interaction.

\begin{figure*}
\includegraphics[clip,scale=0.6]{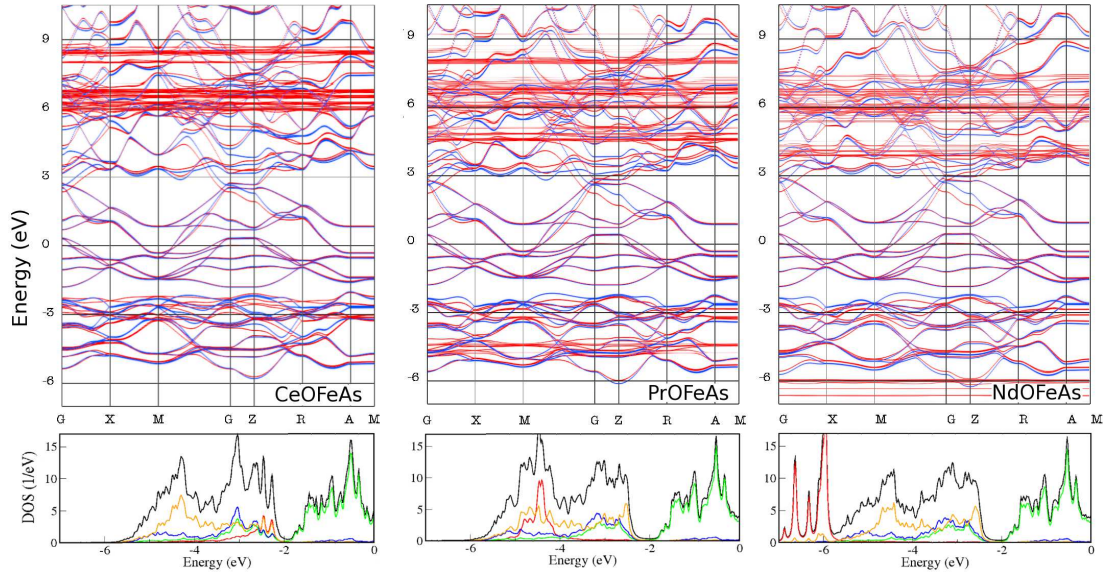}
\caption{\label{ReOFeAs_DOS}
Densities of states obtained within LDA+DMFT for REFeAsO (RE=Ce, Pr, Nd): total, partial RE 4$f$, Fe 3$d$, As $4p$ and O 2$p$ shown
by black, red, green, blue, and orange curves, respectively }
\end{figure*}

In Fig.~\ref{ReOFeAs_DOS} we show the calculated LDA+DMFT band structure and partial densities of states for REFeAsO (RE=Ce,Pr,Nd). For the sake of comparison
the LDA bands computed with the RE 4$f$ treated as core are also shown (in blue color) in Fig.~\ref{ReOFeAs_DOS}.
The occupied 4$f$ states are located in the range of energies from -4 to -5 eV in PrOFeAs, and below -6 eV for NdOFeAs.
The occupied 4$f$ states hybridize rather strongly with the oxygen $p$ states, which are located mainly at the top of the As 4$p$/O 2$p$ band. This
4$f$-2$p$ hybridization leads to some modifications at the top of the O/As $p$ band in comparison with the $f$-in-core
band structure. In contrast, the Fe 3$d$ bands
are not modified by the interaction with the 4$f$ states, as one may see in the
upper panel of Fig.~\ref{ReOFeAs_DOS} where blue and red bands corresponding to Fe (close to the Fermi level)
essentially coincide.
The Pr and Nd 4$f$ bands are located well below the Fe 3$d$ bands. This implies that the Pr and Nd
4f shell simply leads to local moment behaviour, quite decoupled from other electronic states.

In contrast, in the Ce compound the occupied 4$f$ band is located in the range of
energies -2.2 to 3 eV, with the highest peaks of the 4$f$ partial DOS at -2.3 and -2.5 eV.
In experimental photoemission measurements of CeFeAsF$_x$O$_{1-x}$ \cite{bondino} a peak
at -1.7 eV has been assigned to the occupied Ce 4$f$ band. Hence,
the Ce 4$f$ occupied states are located just below the Fe 3$d$ bands. Therefore,
in analogy to the case of CeFePO \cite{pourovskii-epl} where the Kondo screening of the
Ce 4$f$ local moment was observed experimentally \cite{bruning}, one may expect Kondo screening
of the 4$f$ local moment by conduction electrons in CeFeAsO, particularly under applied pressure.
In Ref.~\cite{pourovskii-epl} the corresponding Kondo
temperature $T_K$ has been estimated and its evolution under applied pressure investigated.
The predicted rapid increase of $T_K$ as function of pressure
points to a possible competition between superconducting and heavy-fermion phases
in CeFeAsO. This competition may be at origin of the rather
rapid suppression of superconductivity observed in the doped
CeFeAsO under pressure.

\section{Conclusion}

We have estimated the strength of the local Coulomb interaction $U$ in the Fe 3$d$
and RE 4$f$ shells of the REFeAsO compounds. For both the 3$d$ and Ce 4$f$ states
the $U$ values in REFeAsO are substantially larger than the corresponding values for elemental
Fe and Ce metals. This can be explained by the layered structure of these materials resulting
in less effective screening of the local Coulomb interactions.
The obtained value of U for the Fe 3$d$ shell is equal
to 3.69 eV, which is rather close to the value of $U$ used in recent LDA+DMFT calculations \cite{haule}.
The value of $U$ for the Ce 4$f$ shell obtained by constrained LDA calculations for CeFeAsO is above 9 eV.
We investigated correlation effects on the Re 4f shell of the REFeAsO compounds (RE=Ce,Pr,Nd) using the LDA+DMFT
approach in conjunction with an atomic approximation.
In the Pr and Nd compound the occupied 4$f$ band is located well below the Fe 3$d$ bands, implying
unscreened local moment behaviour of the 4$f$ states.
In CeFeAsO the occupied 4$f$ band is located just below the Fe 3$d$ band, which may lead to
Kondo screening under applied pressure and points to a competition between superconducting and
heavy-fermion phases in this compound.

\acknowledgements
We thank M.Imada and K.Nakamura for useful discussions.



\begin{thebibliography}{1}
\bibitem{Kamihara-LOFA}Y. Kamihara {\it et al.}, J. Am. Chem. Soc. \textbf{130}, 3296 (2008).
\bibitem{cruz} C. de la Cruz {\it et al.}, Nature \textbf{453}, 899 (2008).
\bibitem{struct} Zhi-An Ren {\it et al.}, Europhys. Lett. \textbf{83}, 17002 (2008); C. H. Lee {\it al.}, arXiv:0806.3821 (2008).
\bibitem{zhao} Zhao J. {\it et al.}, arXiv: 0806.2528(2008).
\bibitem{slezak} J. A. Slezak {\it et al.}, PNAS \textbf{105}, 3208 (2008).
\bibitem{malaeb} Walid Malaeb {\it et al.}, arXiv:0806.3860 (2008).
\bibitem{ding} H. Ding {\it et al.}, arXiv:0807.0419 (2008).
\bibitem{photo} E. Z. Kurmaev {\it et al.}, arXiv:0805.0668(2008); T. Kroll {\it et al.}, arXiv 0806.2625 (2008); F. Bondino {\it et al.}, arXiv: 0807.3781 (2008).
\bibitem{haule} K. Haule, J.H. Shim, and G. Kotliar, Phys. Rev. Lett. \textbf{100}, 226402 (2008).
\bibitem{mazin2} I. I. Mazin {\it et al.}, arXiv: 0806.1869 (2008).
\bibitem{pickett} Z. P. Yin {\it et al.}, arXiv:0804.3355 (2008).
\bibitem{dong} J. Dong {\it et al.}, Europhys. Lett \textbf{83}, 27006 (2008).
\bibitem{boris}  A.V. Boris {\it et al.}, arXiv: 0806.1732 (2008).
\bibitem{ren3} Zhi-An Ren {\it et al.}, Europhys. Lett. \textbf{83}, 17002 (2008).
\bibitem{wien2k} P. Blaha {\it et al.}, WIEN2K, An augmented planewave+local orbitals program for calculating
crystal properties (Technische Universitat Wien, 2002, Austria); http://www.wien2k.at
\bibitem{aryasetiawan_prb04}  F. Aryasetiawan {\it et al.}, Phys. Rev. B \textbf{70}, 195104 (2004); T. Miyake and F. Aryasetiawan, Phys. Rev. B \textbf{77}, 085122 (2008).
\bibitem{vildosola} Vildosola V. {\it et al.}, Phys. Rev. B {\it in press}, arXiv: 0806.3285 (2008).
\bibitem{nakamura} K.  Nakamura, R. Arita, and M. Imada, arXiv: 0806.4750 (2008).
\bibitem{footnote} Improvement in the pseudo-potential treatment used in \cite{nakamura} may
lead to better agreement between the estimate of these authors, and our values (based on a
full-potential LMTO implementation).
\bibitem{POFA} P.Quebe, L. J. Terb\"{o}chte, and W. Jeitschko, J. Alloys and Compounds \textbf{302}, 70 (2000).
\bibitem{SOFA} N. D. Zhigadlo {\it et al.}, arXiv: 0806.0337 (2008).
\bibitem{dmft}  A. Georges {\it et al.}, Rev. Mod. Phys. \textbf{68}, 13 (1996);V. I. Anisimov {\it et al.}, J. Phys.: Condens. Matter \textbf{9}, 7359(1997); A. I. Lichtenstein, and M. I. Katsnelson, Phys. Rev. B \textbf{57}, 6884 (1998); G. Kotliar {\it et al.}, Rev. Mod. Phys. \textbf{78}, 865 (2006).
\bibitem{pour2007} L. V. Pourovskii \emph{et al.},   Phys. Rev. B \textbf{76}, 235101 (2007).
\bibitem{hub1} J. Hubbard, Proc. R. Soc. London, Ser. A \textbf{276}, 238 (1963).
\bibitem{carnal} W. T. Carnal \emph{et al.},  J. Chem. Phys. \textbf{90}, 3443 (1989).
\bibitem{bondino} F. Bondino {\it et al.}, arXiv:0807.3781 (2008).
\bibitem{pourovskii-epl} L. Pourovskii {\it et al.}, arXiv:0807.1037 (2008).
\bibitem{bruning} E.M. Bruning \emph{et al.}, arXiv:0804.3250(2008).
\end{thebibliography}
\end{document}